\documentclass[preprint2]{aastex}

\shorttitle{Deuteration factory in L1544}
\shortauthors{Vastel et al.}

\begin{document}

\title{The distribution of {\it ortho}--H$_2$D$^+$(1$_{1,0}$--1$_{1,1}$) 
in L1544: tracing the deuteration factory in prestellar cores}

\author{C. Vastel\altaffilmark{1}}
\affil{California Institute of Technology, Mail-code 320-47, Pasadena, CA 91125, USA}
\author{P. Caselli\altaffilmark{2},\altaffilmark{3}}
\affil{INAF - Osservatorio Astrofisico di Arcetri, Largo E. Fermi 5, 50125 Firenze, Italy}
\affil{Harvard-Smithsonian Center for Astrophysics, 60 Garden St., Cambridge, MA 02138, USA}
\author{C. Ceccarelli\altaffilmark{4}}
\affil{Laboratoire d'Astrophysique de l'Observatoire de Grenoble, BP 53, 38041 Grenoble Cedex 9, France}
\author{T. Phillips\altaffilmark{1}}
\author{M.C. Wiedner\altaffilmark{5}}
\affil{I. Physikalisches Institut, Universit\"at zu K\"oln, Z\"ulpicher Str. 77, 50937 K\"oln, Germany}
\author{R. Peng\altaffilmark{6}}
\affil{Caltech Submillimeter Observatory, Hilo, HI, USA}
\author{M. Houde\altaffilmark{7}}
\affil{Department of Physics and Astronomy, University of Western Ontario, London, Ontario N6A 3K7, Canada}
\author{C. Dominik\altaffilmark{8}}
\affil{Sterrenkundig Instituut "Anton Pannekoek'', Kruislaan 403, 1098 Amsterdam, The Netherlands}

\begin{abstract}
Prestellar cores are unique laboratories for studies of the chemical and physical conditions preceding 
star formation. We observed the prestellar core L1544 in the fundamental transition of 
ortho--H$_{2}$D$^{+}$ (1$_{1,0}$--1$_{1,1}$) at different positions over 100$^{\prime\prime}$, and found a strong correlation 
between its abundance and the CO depletion factor. We also present a tentative detection of the fundamental 
transition of para--D$_{2}$H$^{+}$ (1$_{1,0}$--1$_{0,1}$) at the dust emission peak. 
Maps in N$_{2}$H$^{+}$, N$_{2}$D$^{+}$, HCO$^{+}$ and DCO$^{+}$ are used, and 
interpreted with the aid of a spherically symmetric chemical model that predicts the column densities and abundances 
of these species as a function of radius. The correlation between the 
observed deuterium fractionation of H$_{3}$$^{+}$, N$_{2}$H$^{+}$ and HCO$^{+}$ and the observed 
integrated CO depletion factor across the core can be reproduced by this chemical model. In addition a 
simpler model is used to study the H$_{2}$D$^{+}$ ortho--to--para ratio. We conclude that, 
in order to reproduce the observed ortho--H$_{2}$D$^{+}$ observations, the grain radius 
should be larger than 0.3 $\mu$m.

\end{abstract}

\keywords{ISM: molecules --- ISM: individual (L1544) --- radio lines: ISM}

\section{Introduction}

\begin{figure}
\epsscale{.80}
\plotone{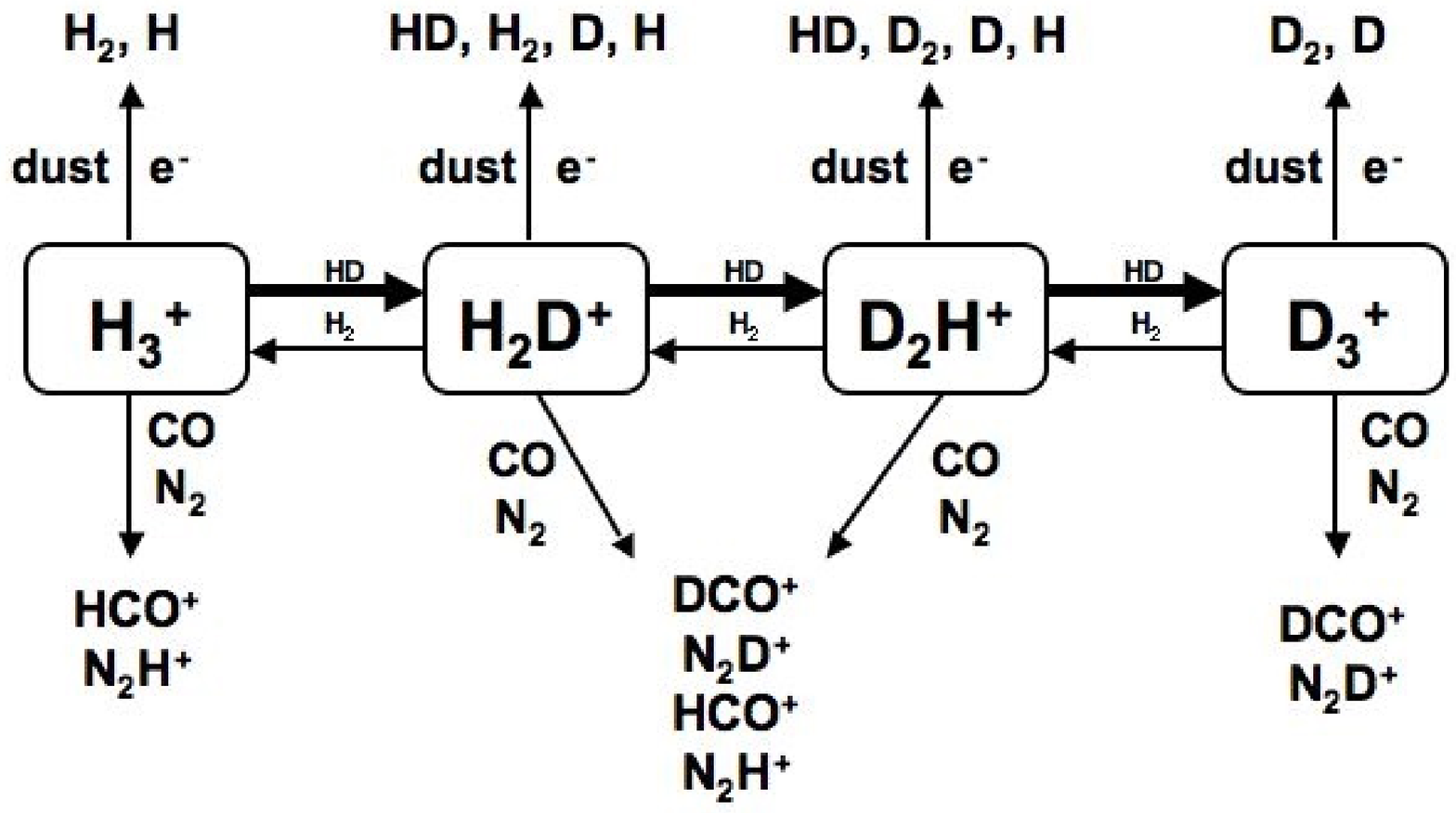}
\caption{Main reactions  involving the deuterated forms of the H$_{3}^{+}$ molecule. When CO and N$_{2}$ are 
depleted, the molecular reactions presented with bold arrows are dominant.\label{reactions}}
\end{figure}

Deuterium bearing species are good probes of the cold phases of molecular clouds prior to star formation. Many recent 
observations point to the fact that their abundances relative to their fully hydrogenated forms are larger, by factors up to 
10$^{5}$, than the solar neighborhood value of $\sim$ 1.5 $\times$ 10$^{-5}$ found by \citet{linsky03}. 
The deuterium fractionation has been evaluated in prestellar cores and low mass protostars from observations of HCO$^+$ and 
N$_2$H$^+$ (Butner et al. 1995; Williams et al. 1998; Caselli et al. 2002a; Crapsi et al. 2004, Crapsi et al. 2005), 
H$_2$CO (Loinard et al. 2001; Bacmann et al. 2003), H$_2$S (Vastel et al. 2003), HNC (Hirota et al. 2003), 
CH$_3$OH (Parise et al. 2004), and NH$_3$ (Roueff et al. 2000, Tin\'e et al. 2000). 
The chemical fractionation process in the gas--phase mainly arises from the difference between 
the  zero-point energies of H$_2$ and HD. Almost incredibly, this 
can lead to a detectable quantity of triply deuterated molecules like ND$_3$ (Lis et al. 2002; 
van der Tak et al. 2002) and CD$_3$OH (Parise et al. 2004). The latter is thought to 
be formed mainly on dust grain surfaces (Charnley et al. 1997) in 
regions where the gas--phase [D]/[H] ratio is enhanced to values larger than 
$\sim$0.1 (Caselli et al. 2002c), as in the cold cores (see below). 
In molecular clouds, hydrogen and deuterium are predominantly in the
form of H$_2$ and HD respectively. So the HD/H$_2$ ratio should closely
equal the D/H ratio. Since the zero-point energies of HD and H$_2$
differ by $\sim$ 410 K, the chemical
fractionation will favor the production of HD compared to H$_2$.
In the dense, cold regions of the
interstellar medium (T $\sim$ 10 K), D will be initially nearly all absorbed into HD. The abundant ion 
available for interaction is H$_3^+$, which gives H$_2$D$^+$: 
\begin{equation}
{\rm H_3^+ + HD  \longleftrightarrow H_2D^+ + H_2 + 230 K}
\end{equation}
The reverse reaction does not occur efficiently in the cold dense clouds where low--mass stars form, 
and where the kinetic temperature is always below 25~K, the ``critical'' temperature above which reaction 
(1) starts to proceed from right to left and limits the deuteration. Therefore, the degree of fractionation of 
H$_2$D$^+$ becomes non-negligible. This primary fractionation can then give rise to other 
fractionations and form D$_2$H$^+$ and D$_3^+$ as first suggested by \cite*{phillips03}:
\begin{equation}
{\rm H_2D^+ + HD  \longleftrightarrow D_2H^+ + H_2 + 180 K}
\end{equation}
\begin{equation}
{\rm D_2H^+ + HD  \longleftrightarrow D_3^+ + H_2 + 230 K}
\end{equation}
We present in Figure \ref{reactions} the main reactions involving these molecules. Note that the effect of theÊ
recombination of H$_3^+$ with electrons in the gas is negligible because of the low electron density in such regions. 
However, the effect of recombination with electrons on negatively charged grain surfaces becomes important when 
depletion increases (cf. Walmsley et al. 2004). 
The dissociation of the deuterated forms of H$_3^+$ is then responsible for the enhancement in the 
[D]/[H] ratio. One specific parameter can enhance this process: the depletion of neutral species (in particular, 
the abundant CO) from the gas-phase (cf. Dalgarno \& Lepp 1984).
In fact, the removal of species which would normally destroy H$_3^+$ \citep[e.g. CO;][]{roberts00} means
that H$_3^+$ is more likely to react with HD and produce H$_2$D$^+$, D$_2$H$^+$ and D$_3^+$. The first model 
including D$_{2}$H$^{+}$ and D$_{3}^{+}$ \citep{roberts03} predicted that these molecules should be as 
abundant as H$_{2}$D$^{+}$ \citep[see also][]{flower04}.

Gas phase species are expected to be depleted at the center of cold, dark clouds, since they tend to stick onto the 
dust grains. A series of recent observations has shown that, in some cases, the
abundance of molecules like CO decreases toward the core center of
cold ($\le$ 10 K), dense ($\ge$ a few 10$^4$ cm$^{-3}$) clouds. L1498: \cite{willacy98}, Tafalla et al. (2002, 2004); 
IC 5146: \cite{kramer99}, \cite{bergin01}; L977: \cite{alves99}; L1544: \cite{caselli99}, Tafalla et al. (2002); 
L1689B: \cite{jessop01}; \cite{bacmann02}; Redman et al. (2002); B68: Bergin et al. (2002); 
L1517B: Tafalla et al. (2002, 2004); L1512: Lee et al. (2003); Oph D: Bacmann et al. (2003), Crapsi et al. (2005); 
L1521F: Crapsi et al. (2004); L183 (L134N): Pagani et al. (2005). These decreases in abundance have been 
interpreted as resulting from the depletion of molecules onto dust grains \citep[see, e.g.,][]{bergin97,charnley97}. 
It is now clear that these drops in abundance are typical of the majority of dense cores (see Tafalla \& Santiago 2004 
for the case of L1521E, the first starless core to be found with no molecular depletion). 

In one of the most heavily CO depleted prestellar cores, L1544,  \cite*{caselli03} detected a strong (brightness 
temperature $\sim$ 1~K) ortho-H$_2$D$^+$(1$_{01}$-1$_{11}$) line, and concluded that H$_2$D$^+$ is one of 
the main molecular ions in the central region of this core. Encouraged by laboratory measurements \citep{hirao03}, 
\citet{vastel04} detected  the para--D$_{2}$H$^{+}$ molecule in its ground transition at 692 GHz. 
They found that, in the prestellar core 16293E, 
D$_{2}$H$^{+}$ is as abundant as H$_{2}$D$^{+}$. These studies supported chemical modelling and the inclusion 
of multiply deuterated species (Roberts et al. 2003; Walmsley et al. 2004; Roberts et al. 2004; Flower et al. 2005; 
Aikawa et al. 2005). 
It appears that in dark clouds affected by molecular depletion, the deuterated forms of the molecular ion H$_3^+$ 
are unique tracers of the core nucleus, the future stellar cradle.  Thus, their study becomes fundamental to the unveiling 
of the initial conditions of the process of star formation (kinematics, physical and chemical structure of pre--stellar cores).

In this paper, we present new observations of the H$_2$D$^+$ (1$_{10}$-1$_{11}$) line toward L1544, 
mapped over $\sim$ 100$^{\prime\prime}$ of the dust peak emission, as well as a tentative detection of the 
D$_{2}$H$^{+}$ (1$_{10}$-1$_{01}$) line. \cite*{caselli03} roughly estimated the size of the H$_2$D$^+$ 
emitting region in this pre--stellar core and suggested a radius of about 3000 AU. But this was based on a 
five-point map and cannot put stringent constraints on the chemical structure. Also a parallel study has 
been done by van der Tak et al. (2005)  with the analysis of the line shape profile. 
Here, the H$_2$D$^+$ map is also 
compared with other high density tracer maps. Due to the poor atmospheric transmission at the frequency of the 
para--D$_{2}$H$^{+}$ fundamental line, this study is limited to the ortho--H$_{2}$D$^{+}$ fundamental line.  
We will present, in the last section, the perspectives on this work that will be opened up by future observatories.

\section{Observations}

The observations were carried out at the Caltech Submillimeter Observatory (CSO), 
between November 2003 and February 2005, under good weather conditions (225 GHz zenith opacity 
always less than 0.06) where the atmospheric transmission is about 30\% at 372 GHz and less than 
 20\% at 692 GHz. Scans were taken toward the peak of the 1.3 mm continuum dust emission of L1544 
 ($\alpha_{2000}~=~05^h~04^m~17^s.21$, $\delta_{2000}~=~+25^\circ~10^\prime~42.8^{\prime \prime}$) 
using the chopping secondary with a throw of 3$^{\prime}$.
The 345 GHz (respectively 650 GHz) sidecab receiver with a 50 MHz acousto-optical spectrometer backend 
was used for all observations with a velocity resolution of 0.06~km~s$^{-1}$ 
(respectively 0.03~km~s$^{-1}$) i.e. $\sim$ 1.6 channels. At the observed frequencies 
of 372.421385(10) GHz for the H$_2$D$^+$ (1$_{10}$-1$_{11}$) and 691.660483(20) 
for the D$_{2}$H$^{+}$ (1$_{10}$-1$_{01}$) lines \citep{amano05}, the CSO 10.4 meters 
antenna has a HPBW of about 20$^{\prime\prime}$ and 11$^{\prime\prime}$ respectively.
We mapped the area in H$_{2}$D$^{+}$ around the dust peak position with a grid spacing of 
20$^{\prime\prime}$ and used the value at the peak from \cite*{caselli03} and integrated 
longer. 
The beam efficiency at 372 GHz (respectively 692 GHz) was measured on Venus, Saturn and Jupiter and 
found to be $\sim$ 60\% (respectively $\sim$ 40\%). Pointing was monitored every 1 hour and half and found to be 
better than 3$^{\prime\prime}$. In the case of H$_{2}$D$^{+}$, the emission is extended compared 
to the beam size of CSO: the efficiency is then about 70\%. If the emission in D$_{2}$H$^{+}$ is extended 
compared to the beamsize of 11$^{\prime\prime}$, then the efficiency at 692 GHz is about 60\%.
The data reduction was performed using the CLASS program of the GAG software developed at IRAM and the 
Observatoire de Grenoble.  \\

\begin{figure}
\epsscale{.80}
\plotone{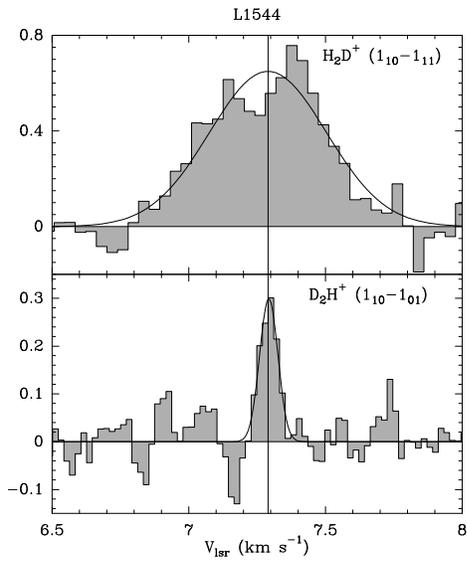}
\caption{Ortho--H$_{2}$D$^{+}$ and para--D$_{2}$H$^{+}$ observations at the dust peak emission corresponding to the 
 (0$^{\prime\prime}$,0$^{\prime\prime}$) position. The temperatures are  T$_{a}^{*}$ in Kelvins. The vertical 
 solid line corresponds to the velocity center of the D$_{2}$H$^{+}$ gaussian fit (7.29 km~s$^{-1}$). \label{lines}}
\end{figure}

Figure \ref{lines} shows the H$_2$D$^+$ and D$_{2}$H$^{+}$ spectra observed toward the dust peak position. 
A Gaussian fit to the H$_2$D$^+$ line gives a LSR velocity of $\sim$7.3 km~s$^{-1}$, but two peaks 
are clearly visible, with velocities 7.1 and 7.4 km~s$^{-1}$ (van der Tak et al. 2005), as also seen in other tracers 
(Tafalla et al. 1998; Caselli et al. 2002a).  This central position has been originally observed by Caselli et al. (2003) 
and studied in detail by van der Tak et al. (2005); here, we improved the sensitivity and used the new value of the 
ortho-H$_{2}$D$^{+}$(1$_{10}$-1$_{11}$) line frequency, recently  measured by Amano \& Hirao (2005).
The on-source integration time for D$_{2}$H$^{+}$ is about 230 min. The D$_{2}$H$^{+}$ feature can 
be fitted with a gaussian with T$_{a}$$^{*}$=0.30 $\pm$ 0.07 K, $\Delta$v=0.08 $\pm$ 0.04km~s$^{-1}$ and 
V$_{LSR}$=7.29 $\pm$ 0.03 km~s$^{-1}$. The solid vertical line corresponds to the velocity from this gaussian fit. 
It is consistent with the central velocity of the H$_{2}$D$^{+}$ line of 7.28 $\pm$ 0.06 km~s$^{-1}$.
However, it is difficult to believe that the observed linewidth represents the real linewidth of the transition as the 
expected thermal linewidth is about 0.25 km~s$^{-1}$ for a kinetic temperature of 7 K, as predicted by dust temperature 
measurement (Evans et al. 2001; Zucconi et al. 2001). This is about 3 times larger than the observed linewidth. 
Formaldehyde observations of \citet{bacmann02} suggest larger gas temperatures (up to 9 K), but H$_2$CO is 
likely frozen in the core center, so that the measured temperature probably reflects the warmer core envelope (Young 
et al. 2004). At high densities (higher than $\sim$ 3 10$^{4}$ cm$^{-3}$), Young et al. (2004) find the gas and dust 
temperature to be between  $\sim$ 7 and 9~K, consistent with what was found by Tafalla et al. (2002) using ammonia. We 
will use in the following a temperature of 8 K in this cloud.
The signal to noise ratio of our D$_{2}$H$^{+}$ observations is not sufficient to get constraints on the kinematics of
the source and the fitted linewidth probably does not represent the real linewidth.  
%One possibility to reproduce the 
%present observations is to consider the D$_2$H$^+$ emission as arising from two layers with slightly different velocity 
%(one at 7.2 km s$^{-1}$ and the other at 7.1 km s$^{-1}$), with the blueshifted component arising from a thick 
%($\tau$ $\sim$ 10) and cold (T$_{\rm kin}$ $\sim$ 4 K) region. 
In absence of a possible explanation for 
the narrow linewidth for the D$_{2}$H$^{+}$ profile, we will consider, 
in the following, that we only have a tentative detection, and will then use an upper limit.

\begin{figure}
\epsscale{.80}
\plotone{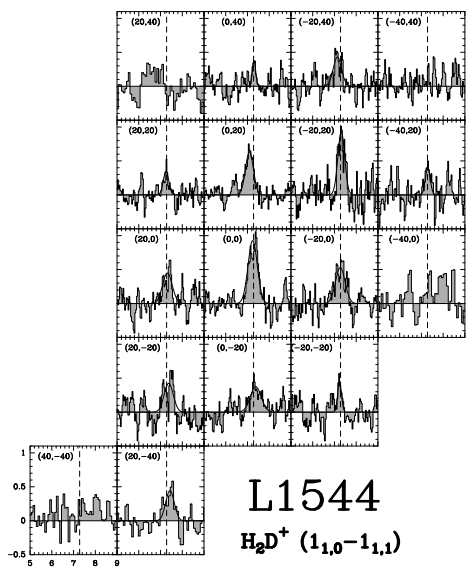}
\caption{Map of the H$_2$D$^+$ (1$_{10}$-1$_{11}$) line centered on the dust peak of L1544. 
The Y-axis represents the main beam temperature. The position is indicated in arcseconds in right 
ascension and declination offsets from the central position, on the top left of each spectrum. 
The reference for the velocity center at the (0$^{\prime\prime}$,0$^{\prime\prime}$) 
position (7.28 km~s$^{-1}$) is indicated in dashed lines. \label{map}}
\end{figure}

Figure \ref{map} shows the H$_2$D$^+$ spectra around the central position (0$^{\prime\prime}$,0$^{\prime\prime}$) 
of the dust peak emission. 
The offset positions shown in the upper left are in arcseconds. A Gaussian fit for each detected line is plotted, using the 
CLASS program, and the line parameters are presented in Table \ref{line_parameters}. A double-peak profile seems to 
appear in the south-east as well as in the central part of the map. 
Considering the rms value, it is only possible to say that this non Gaussian profile is localized in the central 
positions around the dust peak emission. A possible interpretation could be the presence of two different layers 
with different velocities along the line of sight. Another explanation could be that the observed profiles 
are affected by absorption in a low-density (10$^{4}$ cm$^{-3}$) foreground layer redshifted ($\sim$ 0.08 km~s$^{-1}$) 
relative to the high-density core, as found by \citep{williams99} in the
case of N$_2$H$^+$(1--0) mapped at high spatial resolution. A more detailed study of the H$_2$D$^+$ line profile toward 
the L1544 dust peak has been recently carried out by van der Tak et al. (2005), who suggested that the 
observed H$_2$D$^+$ line, besides being broadened by the central 
infall,  can also be absorbed in the outer parts of the core. The presence of a central dip in the H$_2$D$^+$ profile of at 
least four spectra across the L1544 map (see Fig.~\ref{map}) favours this scenario.\\

\begin{figure}
\includegraphics[angle=-90,scale=.80]{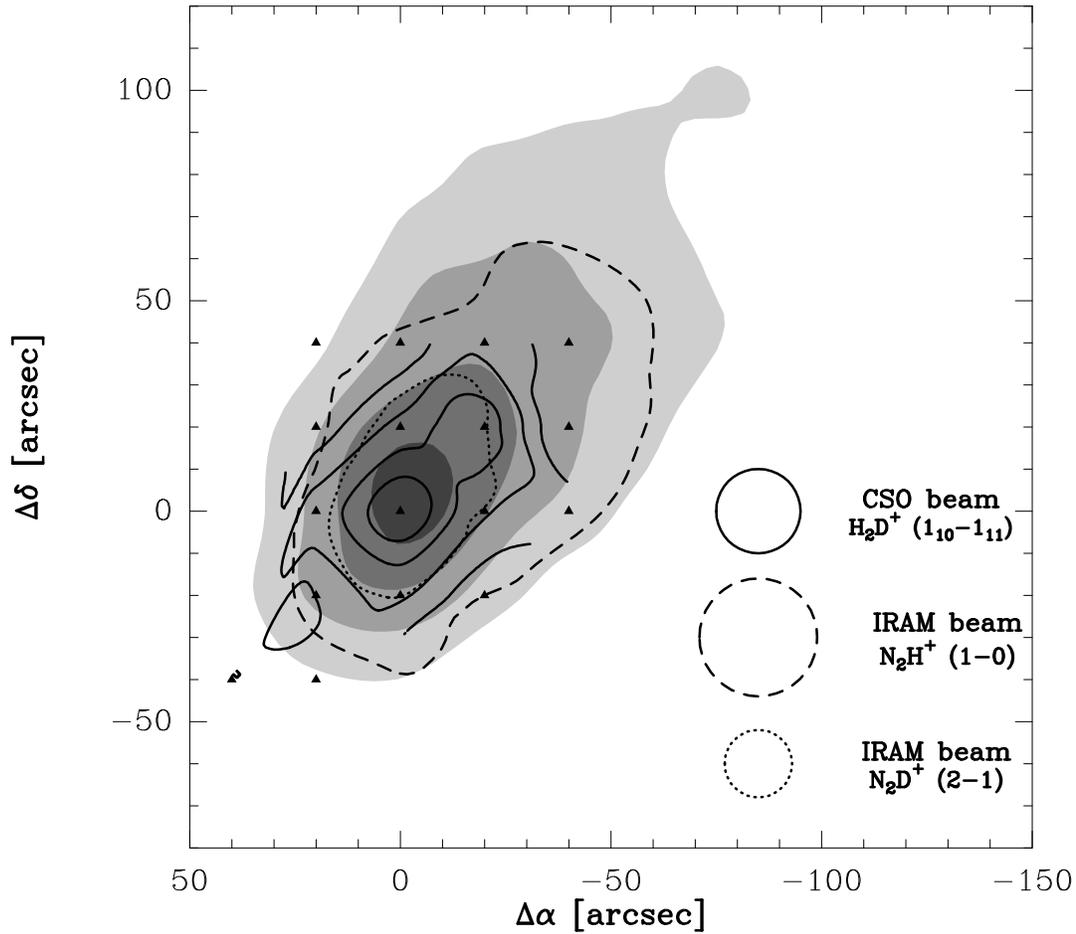}
\caption{Integrated intensity maps of H$_2$D$^+$ (1$_{10}$-1$_{11}$), N$_2$H$^+$ (1-0) and N$_2$D$^+$ 
(2-1) superposed on the 1.3 mm continuum emission map smoothed to a resolution of 22$^{\prime\prime}$ 
(gray scale). Contour levels are 30\%, 50\%, 70\% and 90\% of the peak (0.54 K~km~s$^{-1}$ 
for H$_2$D$^+$), and 50\% of the peak (5.5 K~km~s$^{-1}$ for N$_2$H$^+$ and 2.1 K~km~s$^{-1}$ for N$_2$D$^+$). 
The observed positions in H$_2$D$^+$ are reported as triangles.
The (0$^{\prime\prime}$,0$^{\prime\prime}$) position corresponds to $\alpha_{2000}~=~05^h~04^m~17^s.21$, 
$\delta_{2000}~=~+25^\circ~10^\prime~42.8^{\prime \prime}$. The beam sizes of the observations are also 
reported in the lower right corner of the figure. \label{contour}}
\end{figure}

Figure \ref{contour} shows the integrated intensity map ($\int{}{}T_{mb} dv$) of ortho-H$_2$D$^+$ (1$_{10}$-1$_{11}$), 
together with maps of N$_2$H$^+$ (1-0) and N$_2$D$^+$ (2-1) obtained by Caselli et al. (2002a) and the 1.3~mm continuum 
emission map from Ward--Thompson et al. (1999), smoothed to a resolution of 22$^{\prime\prime}$. We note the close 
similarity between the H$_2$D$^+$ and the N$_2$D$^+$ maps and this will be discussed in the next sections. 
In this paper, we studied the chemistry using the maps made by Caselli et al. (2002a) in H$^{13}$CO$^{+}$, HC$^{18}$O$^{+}$, 
DCO$^{+}$, D$^{13}$CO$^{+}$, C$^{17}$O and C$^{18}$O with the IRAM 30~m telescope. 

\section{Column density and abundance determinations} 
 
The observed molecular ions maps presented in Figure \ref{contour} show a general correlation, despite different 
 beamwidths, with the distribution of dust continuum emission, in contrast to C$^{18}$O (1-0) and 
 C$^{17}$O (1-0) (Caselli et al. 1999), which give clear evidence for depletion of CO at positions 
 close to the continuum peak. H$_2$D$^+$ (1$_{10}$-1$_{11}$), N$_2$D$^+$ (2-1) and to a lesser spatial extent 
 N$_2$H$^+$ (1-0) appear to trace the dust continuum. From these maps N$_2$H$^+$ does not seem to be 
 depleted at the dust peak position. \\
In order to compare the observed species and put constraints on chemical models, we 
need to infer the column densities and abundances of H$_2$D$^+$ and D$_2$H$^+$ (defined as 
$N(i)/N(\rm H_2)$ for a generic species $i$, with $N(\rm H_2)$ derived
from the 1.3~mm dust continuum emission map of Ward--Thompson et al. 1999). 
Assuming LTE conditions, we can estimate the optical depth at the line center 
from the observed line intensities:
 \begin{equation}
{\rm T_{mb} = [J_{\nu}(T_{ex})-J_{\nu}(T_{bg})](1-e^{-\tau})}
\end{equation}
where $J_{\nu}(T) = (h\nu/k)/(e^{h\nu/kT}-1)$ is the radiation temperature 
of a blackbody at a temperature T, and T$_{bg}$ is the cosmic background 
temperature of 2.7 K. The column density is then given by:
\begin{equation}
{\rm N_{tot}=\frac{8\pi\nu^3}{c^3}\frac{Q(T_{ex})}{g_uA_{ul}}\frac{e^{E_{u}/T_{ex}}}{e^{h\nu/kT_{ex}}-1}\int_{}^{}\tau\, dv}
\end{equation}
where Q(T$_{ex}$) is the partition function:
\begin{equation}
{\rm Q(T_{ex})=\sum_{i=0}^{\infty} (2I+1)exp(-E_{i}/kT_{ex})}
\end{equation}
 In the case of the H$_2$D$^+$ transition, g$_u$ = 9, 
A$_{ul}$ = 1.04 10$^{-4}$ s$^{-1}$, E$_{ul}$ = 17.9 K; in the case of the D$_2$H$^+$ transition, g$_u$ = 9, 
A$_{ul}$ = 4.55 10$^{-4}$ s$^{-1}$, E$_{ul}$ = 33.2 K. 
Using equation 4, we can estimate the upper limit on the D$_{2}$H$^{+}$ main beam temperature, under the 
assumption that the conditions of LTE are valid. With an excitation temperature of 8 K, the maximum 
main beam temperature reaches 0.53 K, which is about the observed main beam temperature of the D$_{2}$H$^{+}$ 
tentative detection, in the case where the spatial extent is larger than the beamsize of 11$^{\prime\prime}$. 
The derived column densities depend on the assumed value of the excitation temperature. Within the 7--9 K 
temperature range, the column density at the dust peak can vary by a factor of 2. At larger distance, considering 
the increasing kinetic temperature and the decreasing molecular hydrogen density, the excitation temperature 
could be as low as 6 K. However, at these positions, the H$_{2}$D$^{+}$ column densities should only be decreased 
by a factor $\sim$ 2.5. Consequently we used a constant T$_{ex}$ for our observations as an approximation. 
The corresponding line parameters and column densities are presented in Table \ref{line_parameters}. The upper limit on 
the para--D$_{2}$H$^{+}$ column density has been 
calculated using the thermal linewidth at 8 K (0.27 km~s$^{-1}$).  \\
At the (0$^{\prime\prime}$,0$^{\prime\prime}$) position, the column densities of ortho--H$_2$D$^+$ and para--D$_2$H$^+$ are 
1.8 $\times$ 10$^{13}$~cm$^{-2}$ and $<$2.3 $\times$10$^{13}$ cm$^{-2}$, respectively (see Table 1). The 
abundances of species $i$, $X(i)$, have been determined dividing the column densities $N(i)$ by the associated H$_2$ 
column density derived from the 1.3~mm dust continuum emission.  At the dust peak, we obtain
$X(\rm ortho-H_2D^+)$ $\simeq$ 1.5$\times$10$^{-10}$ and $X(\rm para-D_2H^+)$ $<$ 1.8$\times$10$^{-10}$. 

\begin{figure}
\includegraphics[angle=0,scale=0.6]{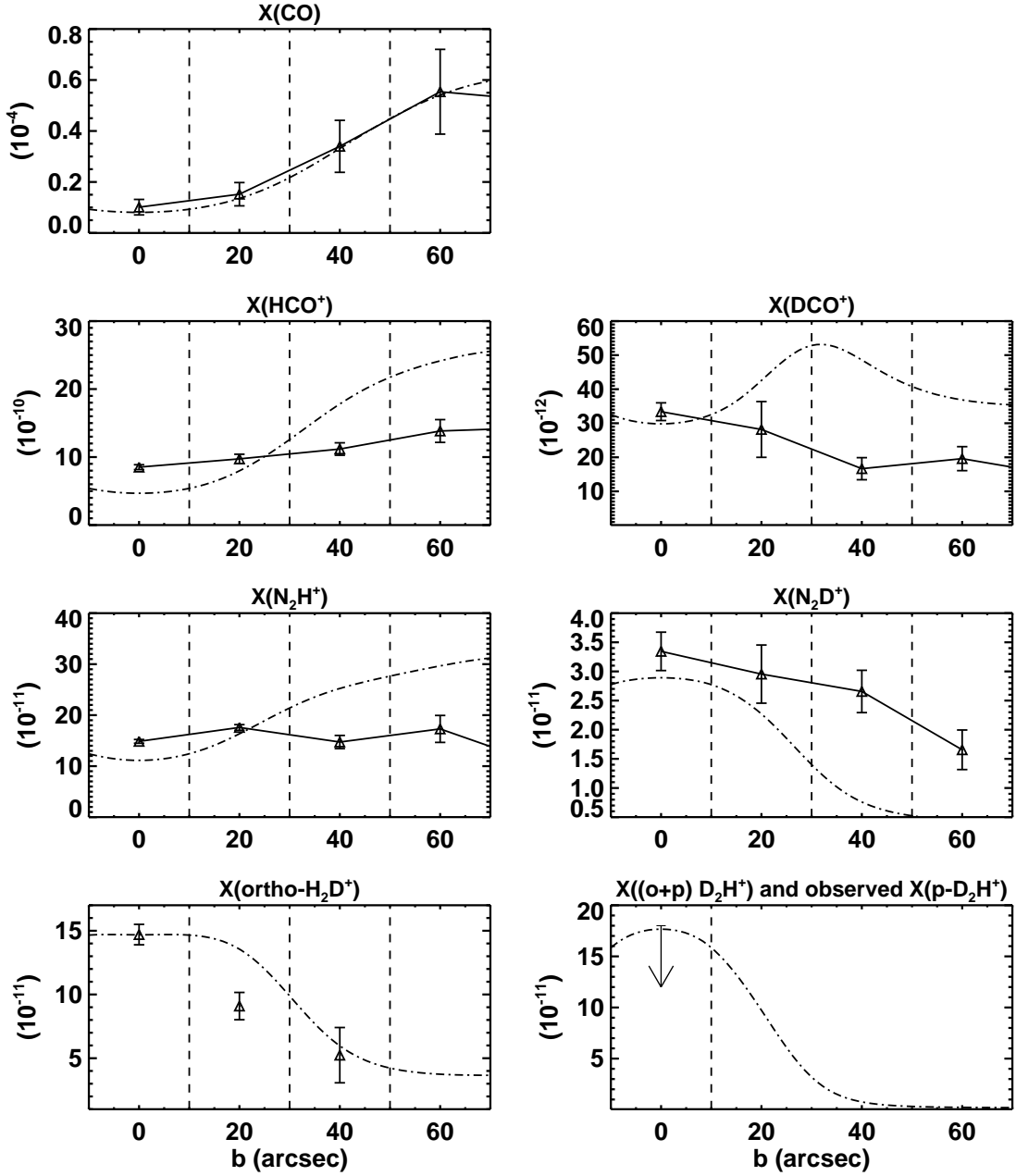}
\caption{Variation of the CO (in a 20" beam), HCO$^{+}$, DCO$^{+}$, N$_{2}$H$^{+}$, N$_{2}$D$^{+}$, 
H$_{2}$D$^{+}$ and D$_{2}$H$^{+}$ abundances as a function of distance. The triangles represent the 
observation points averaged in the bins delimited  by the dashed lines, and the dot-dashed lines represent the results from a "best-fit" 
model. For H$_{2}$D$^{+}$, note that the variation of the observed ortho abundance is compared with the modeled ortho variation 
(see text). Also for D$_{2}$H$^{+}$, the upper limit on the para abundance is reported on the plot of the modeled ortho + para variation. 
\label{abundances}}
\end{figure}

We present in Figure \ref{column_density} the variation of the observed 
ortho--H$_{2}$D$^{+}$, CO, H$_{2}$, HCO$^{+}$, DCO$^{+}$, 
N$_{2}$H$^{+}$, and N$_{2}$D$^{+}$ column densities (crosses) across the 
core (as a function of the impact parameter, the projected distance 
from the dust peak) as well as 
an upper limit on the para--D$_{2}$H$^{+}$. Note that we used the ortho--H$_{2}$D$^{+}$ observations. 
We did not use any ortho--to--para ratio to estimate the total H$_{2}$D$^{+}$ column density or abundance 
because of the large uncertainty on this ratio. A more thorough discussion on the orth--to--para ratio for 
H$_{2}$D$^{+}$ and D$_{2}$H$^{+}$ will be performed in section 4.2.2. 
The observed column densities have then been averaged within the ranges delimited by the vertical dashed lines, at the 
positions (2i + 1) $\times$ 10" where i=0,1,2,3... and are represented by triangles. The same computation 
was performed to present in Figure \ref{abundances} the variation of the abundances as a function of the 
distance to the core center, limited to the central 70$^{\prime\prime}$ (r $\sim$ 9800 AU). 
We will compare in section 4.1 these observations with the result from a best-fit model (dashed lines).\\
 
From Figures \ref{column_density} and \ref{abundances}, we see that only CO is strongly depleted in the core center. 
Note that we used, in the CO column density computation: $^{16}$O/$^{18}$O=560 (Wilson \& Rood 1994), and 
$^{18}$O/$^{17}$O=4 (Wouterloot et al. 2005; Ladd 2004). Defining the CO depletion factor, $f_{\rm D}$, 
as the ratio of the CO ``canonical'' abundance ([CO]/[H$_2$] = 9.5$\times$10$^{-5}$; Frerking, Langer \& Wilson 1982) 
and the observed CO abundance (N(CO)/N(H$_{2}$)), 
Caselli et al. (1999) found $f_{\rm D}$ = 10 toward the dust peak and concluded that the most likely explanation for the 
low CO abundance is the freeze--out of CO onto dust grains at high densities (n $>$ 10$^{5}$ cm$^{-3}$). The 
corresponding radius of the depleted region is r $\sim$ 6500 AU ($\sim$ 45$^{\prime\prime}$).

The CO abundance is a critical parameter in the deuteration of the molecular ion H$_3^+$ (see Figure \ref{reactions}). 
In fact, from the abundance profiles presented in Figure \ref{abundances} it is clear that the degree of deuterium 
enhancement (with DCO$^{+}$, N$_{2}$D$^{+}$, and H$_{2}$D$^{+}$) increases toward the dust peak emission of 
L1544 where CO is highly depleted, as previously found by Caselli et al. (2002b).
N$_{2}$H$^{+}$ does not show any signs of depletion. It is mainly formed by interaction between H$_{3}$$^{+}$ and 
molecular nitrogen, which is likely to be the main repository of nitrogen in the gas phase. N$_{2}$ is only slightly
less volatile than CO (factor of $\sim$0.9; \"Oberg et al. 2005), so that the two neutrals are expected to behave similarly.  
However, N$_2$H$^+$ is destroyed by CO (Bergin et al. 2001, 2002; Pagani et al. 2005; Aikawa et al. 2005), 
so that the CO freeze--out implies a drop in the destruction rate of 
N$_2$H$^+$, which at least partially balance the lower formation rate due 
to the N$_2$ freeze--out (see also Aikawa et al. 2001 for a discussion on 
this point). In fact, N$_2$H$^+$ is observed to survive in the gas phase at
higher densities ($\sim$10$^6$ cm$^{-3}$) compared to CO 
($\sim$10$^5$ cm$^{-3}$). This is also confirmed by the deuterium 
fractionation observed in N$_2$H$^+$ ($\sim$0.2), about 5 times larger 
than that measured in HCO$^+$ (Caselli et al. 2002b). HCO$^+$ is mainly 
formed via H$_3^+$ + CO and destroyed by dissociative recombination,
so that its abundance is simply reduced by the freeze--out of CO, its parent 
species.  From Figure \ref{abundances} it seems that the HCO$^{+}$ abundance 
is reduced at the dust peak, and increases at larger distance.\\

\subsection{Correlations}

\begin{figure}
\includegraphics[angle=0,scale=0.7]{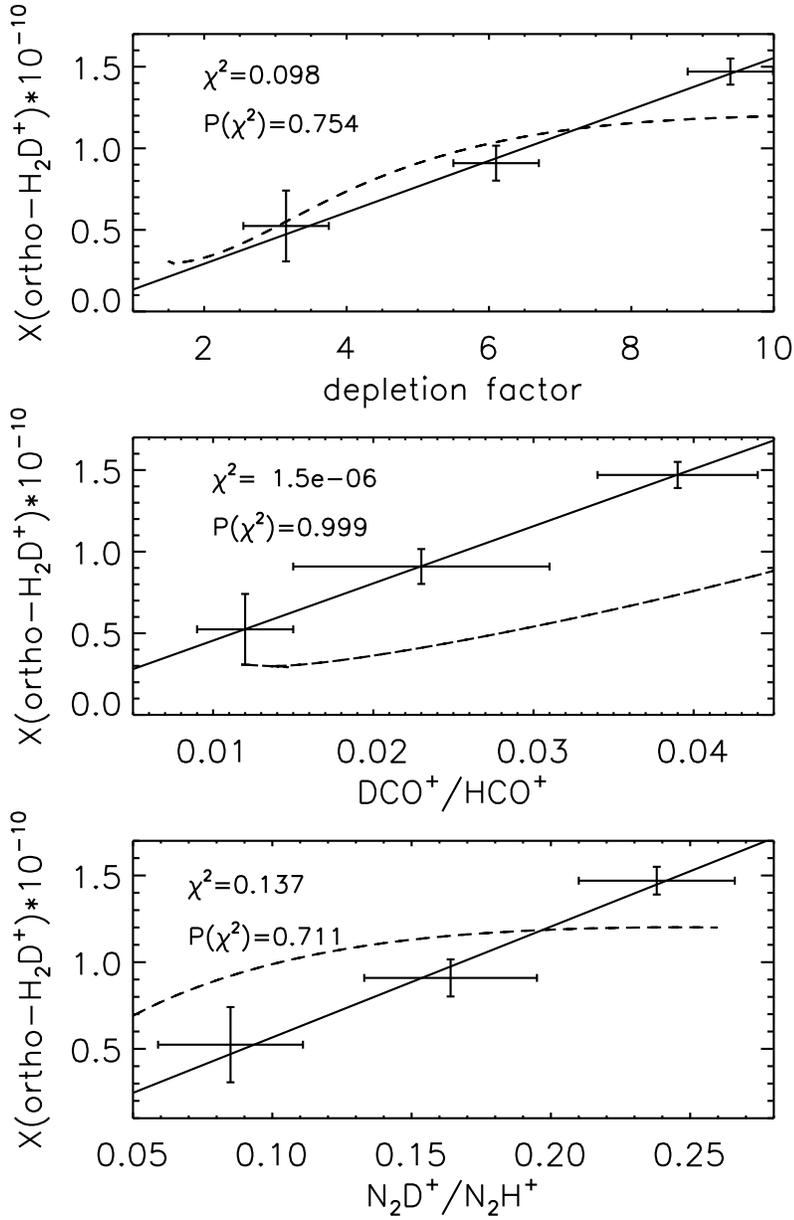}
\caption{Variation of the observed ortho--H$_{2}$D$^{+}$ as a function of the depletion factor, DCO$^{+}$/HCO$^{+}$ ratio 
and N$_{2}$D$^{+}$/N$_{2}$H$^{+}$ ratio. The $\chi^{2}$ parameter and its probability are reported 
when a correlation is established. Superposed are the results from the best-fit model (see section 4.1) \label{correlations}}
\end{figure}

In Figure \ref{correlations}, we show the correlation between the ortho--H$_{2}$D$^{+}$ abundances at the 
0$^{\prime\prime}$, $\pm$ 20$^{\prime\prime}$ and $\pm$ 40$^{\prime\prime}$ distance from the dust peak 
and the CO depletion factor, the DCO$^{+}$/HCO$^{+}$ ratio and the N$_{2}$D$^{+}$/N$_{2}$H$^{+}$ ratio. 
As intuitively expected, the ortho-H$_{2}$D$^{+}$ abundance appears to be well correlated with the CO depletion 
factor (see Figure \ref{reactions}). Also, the degree of deuterium enhancement in the HCO$^{+}$ and 
N$_{2}$H$^{+}$ molecules (measured from the DCO$^{+}$/HCO$^{+}$ and N$_{2}$D$^{+}$/N$_{2}$H$^{+}$ 
ratios) increases linearly with the ortho-H$_{2}$D$^{+}$ abundance. 
The $\chi^{2}$ parameter is calculated for the three points where the impact parameters are 0$^{\prime\prime}$, 
20$^{\prime\prime}$ and 40$^{\prime\prime}$:
\begin{equation}
{\rm \chi^{2}~=~\sum_{i=0}^{2}\left(\frac{X_{obs}(i)-X_{fit}}{\sigma_{X_{obs}(i)}}\right)^{2}}
\end{equation}
where X$_{obs}$ and X$_{fit}$ are the observed and fit values of the abundance respectively, and $\sigma_{X_{obs}}$ 
is the uncertainty in X$_{obs}$. The associated probabilities are reported when a correlation is established. 
The surprisingly high confidence level for the correlations between H$_{2}$D$^{+}$ and the degree of 
deuteration in the HCO$^{+}$ and N$_{2}$H$^{+}$ molecules ($\sim 100\% $), confirms that 
H$_{2}$D$^{+}$ dominates the fractionation of these molecules at low temperatures.

\section{Chemical modelling}

In this section we will interpret the observations using chemical models. We will adopt a two-way strategy. 
First (in section 4.1), we use a full chemical model applied to a density structure derived from continuum 
observations to produce an overall fit to all line observations presented in the previous section. In a second step 
(section 4.2), we 
use a simplified chemical model focussing on the chemistry of H$_3^+$ deuteration, in order to better 
understand the relation between CO depletion and deuteration, and even to provide some estimates of the 
ortho--to--para ratio in the deuterated forms of H$_3^+$ that can be derived from our observations.
  
\subsection{The ``best--fit'' model}

\begin{figure}
\includegraphics[angle=0,scale=0.8]{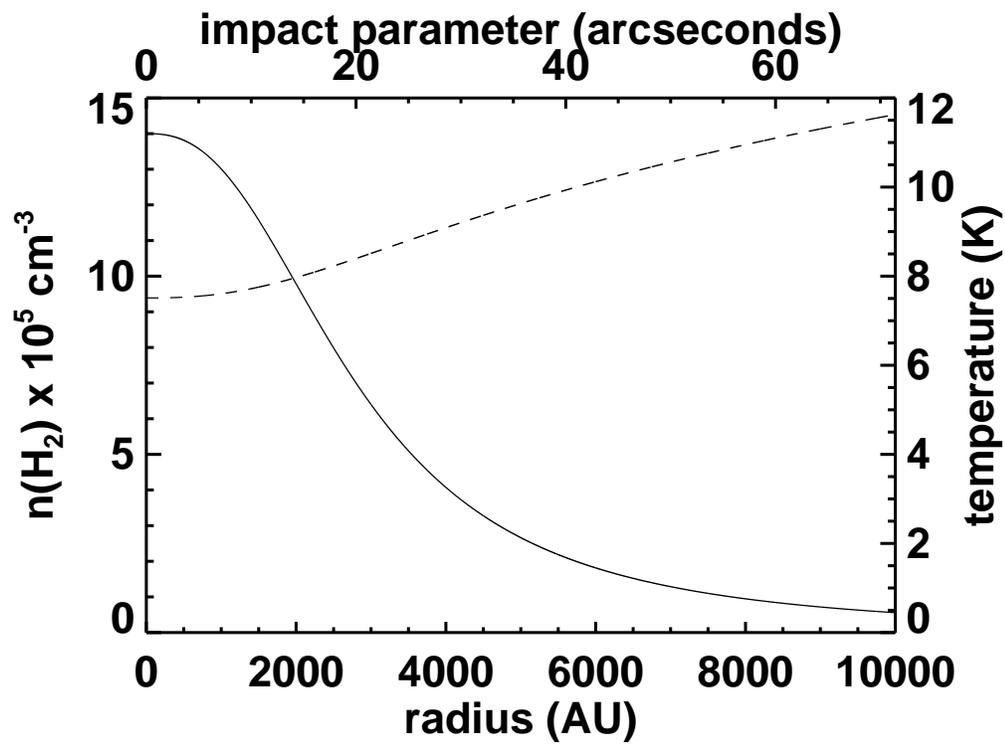}
\caption{Density (plain line) and temperature (dashed line) profiles as a function of the radius (in astronomical units) 
or the impact parameter (in arcseconds), used for the best-fit model.\label{profiles}}
\end{figure}

\begin{figure}
\includegraphics[angle=0,scale=0.6]{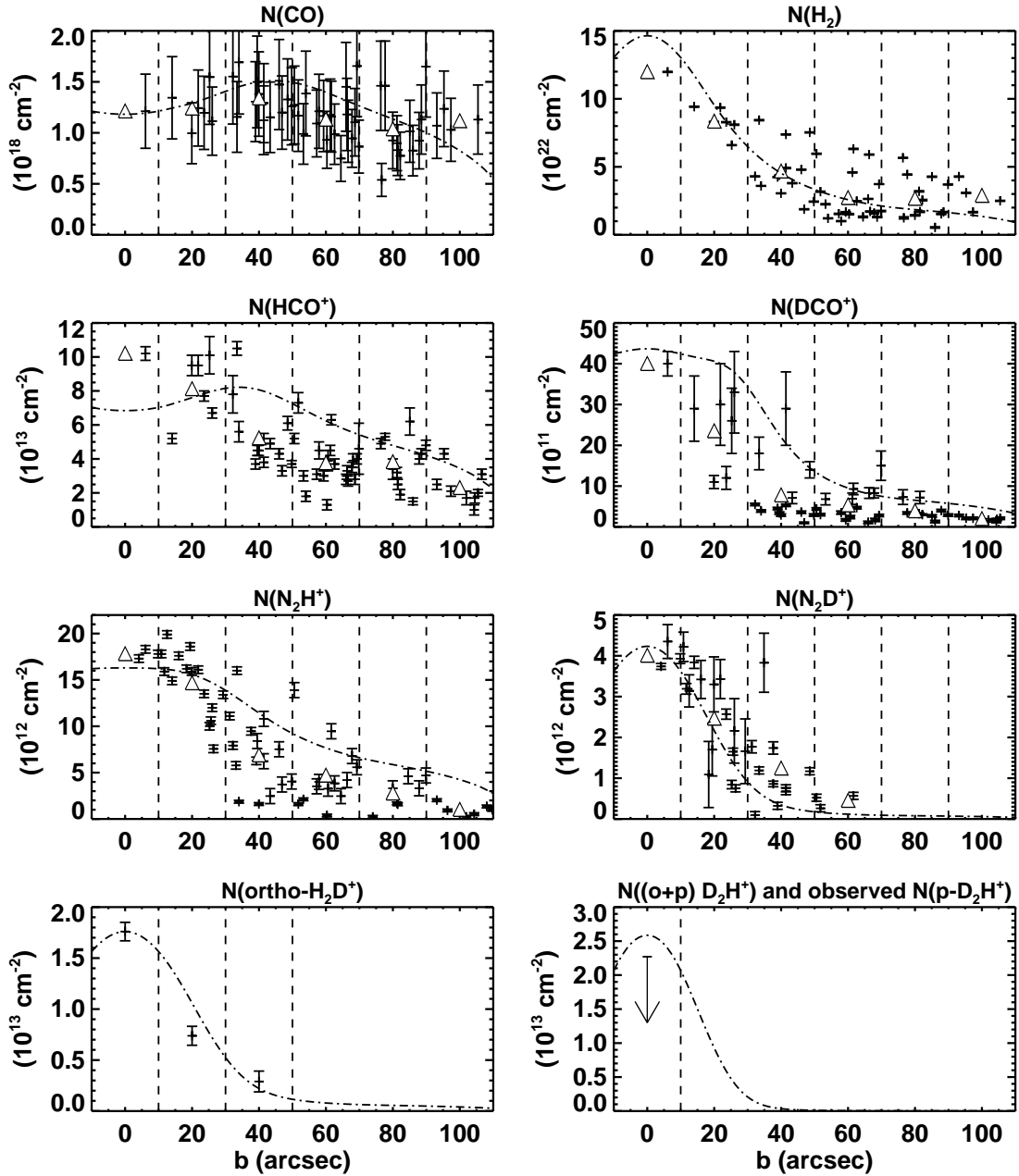}
\caption{Variation of the CO (in a 20" beam), H$_2$, HCO$^{+}$, DCO$^{+}$, N$_{2}$H$^{+}$, N$_{2}$D$^{+}$, 
H$_{2}$D$^{+}$ and D$_{2}$H$^{+}$ column densities as a function of distance. The cross represent the 
observation points and the 3 $\sigma$ error, the triangles represent the observation points averaged in the bins delimited by 
the dashed lines, and the dot-dashed lines represent the results from a "best-fit" model. For H$_{2}$D$^{+}$, note that the 
variation of the observed ortho column density is compared with the modeled ortho variation (see text). Also for D$_{2}$H$^{+}$, the
upper limit on the para column density is reported on the plot of the modeled ortho + para variation.\label{column_density}}
\end{figure}

To more quantitatively understand the column density and abundance correlations shown in Figures 
\ref{column_density}, \ref{abundances} we used the model described in Caselli et al. (2002b), 
updated so that it now includes the multiply deuterated
forms of H$_3^+$ (as in Crapsi et al. 2005) and new values of the binding energies of CO and N$_2$, following 
the measurements by \"Oberg et al. (2005) as well as other modifications to better account for the physical 
structure of the core.  Briefly, the model considers a spherically symmetric cloud, with the density gradient
as derived by Tafalla et al. (2002), using the 1.3~mm dust continuum emission data from Ward--Thompson et al. 
(1999), where the central density is $n({\rm H_2})$ = 1.4$\times$10$^6$ cm$^{-3}$ (see Figure \ref{profiles}). 
The temperature profile has been included, using the recent findings of Young et al. (2004), where the temperature 
is about 7.5~K at the center and reaches about 12~K at the edge (see Figure \ref{profiles}). 
The chemical network contains CO, O and N$_2$, which 
can freeze--out onto dust grains and desorb due to cosmic--ray impulsive heating (as in Hasegawa and Herbst 1993). 
The initial abundances are: $X_i$(CO)=9.5$\times$10$^{-5}$ (cf. Frerking et al. 1982), 
$X_i$(N$_2$)=4.0$\times$10$^{-5}$ (slightly smaller than 6.6$\times$10$^{-5}$, the cosmic value from 
Snow \& Witt 1996, assuming that all nitrogen is in molecular form), 
and $X_i$(O)=$X_i$(CO)/2, a factor of two less than what is typically found 
in gas--phase--only chemical models (e.g. Lee et al. 1996).
The abundances of the molecular ions (N$_2$H$^+$, H$_3$O$^+$ and HCO$^+$ and their deuterated forms) are 
calculated in terms of the instantaneous abundances of neutral species, assuming that the timescale for ion chemistry 
is much shorter than that for freeze--out. The electron fraction has been computed, as described in Caselli et al. 
(2002b), using  a simplified version of the reaction scheme of Umebayashi \& Nakano (1990), where the chemistry 
of a generic molecular ion "mH+" is taken into account, assuming formation due to proton transfer with H and 
destruction by dissociative recombination with electrons and recombination on grain surfaces (using rates from 
Draine \& Sutin 1987). The ``MRN'' grain size distribution has been used \citep{mathis77}.
We adopted the rate coefficients for the proton--deuteron exchange reactions recently measured 
by Gerlich et al. (2002).  The model is run until the column density of C$^{17}$O toward the core center reaches 
the observed value (N(C$^{17}$O) = 6$\times$10$^{14}$ cm$^{-2}$; Caselli et al. 2002b). 

The ``best--fit'' parameters of the model, which best reproduce the observed molecular column densities at the dust 
peak and the observed column density and abundance profiles, are the following: 
\begin{itemize}
\item a cosmic--ray ionization rate of $\zeta$ = 1.3$\times$10$^{-17}$ s$^{-1}$, standard value, typically used in 
chemical models (since Herbst \& Klemperer 1973),
\item binding energies for CO and N$_2$ of $E_{\rm D}$(CO) = 1100 K and $E_{\rm D}$(N$_2$) = 900 K, 
values close to the binding energies measured for CO on water ice (Collings et al. 2003); the ratio 
between N$_2$ and CO binding energies (0.8) being comparable to the value (0.9) recommended by \"Oberg et al. (2005), 
\item a binding energy for atomic oxygen of $E_{\rm D}$(O) = 750 K,  used in current gas--grain chemical models (e.g. Aikawa et al. 2005), 
\item a minimum size of the dust grains, in the MRN distribution of $a_{\rm min}$ = 5$\times$10$^{-6}$ cm 
(10 times larger than in MRN),
\item a sticking coefficient of $S$ = 1.0 (Burke \& Hollenbach 1983), 
\item an initial abundance of metals (assumed to freeze--out with a rate similar to that of CO), 
$X$(M$^+$) = 10$^{-6}$.
\end{itemize}

In Figures \ref{column_density}, \ref{abundances}, \ref{correlations} we overlaid 
the results from the best-fit model (in dashed lines), to the observations. For the H$_{2}$D$^{+}$ plots, we present 
the ortho-H$_{2}$D$^{+}$ observation points, and scaled the total-H$_{2}$D$^{+}$ result from the model by 2.3, 
the factor between the predicted and the observed value at the dust peak position (assumed constant across the core). 
From this, an ortho--to--para ratio of $\sim$ 0.8 can be deduced for H$_2$D$^+$, but, considering the uncertainties 
associated with this parameter (e.g. Pagani et al. 1992, Flower et al. 2005), we will postpone a discussion on this topic
in the following section, where a parameter space exploration will be presented.
For the D$_{2}$H$^{+}$ plot, we present the upper limit on the observed para--D$_{2}$H$^{+}$ compared with the 
(total) D$_{2}$H$^{+}$ result from the model. 
The ortho--H$_{2}$D$^{+}$ column densities and abundances observed at 0", $\pm$ 20" and $\pm$ 40" are 
well reproduced by the model, within the error bars (see Figures \ref{column_density}, \ref{abundances}). 
Note that although 
we need to assume a high degree of CO depletion in order to explain the CO observations, it appears that 
the HCO$^{+}$ column density is slightly under-estimated for the model at the dust peak emission. 
However, the strong variation (factor of 2) seen in Figure \ref{column_density} between 0 and 
15$^{\prime\prime}$ could decrease the central HCO$^{+}$ column density. The discrepancy 
between the HCO$^{+}$ and DCO$^{+}$ column densities at larger distances can 
be explained by the uncertainties on the optical depth, since the less optically thick isotopes (HC$^{18}$O$^{+}$ and 
D$^{13}$CO$^{+}$) have only been used for the central position. Also the beamsize for the HC$^{18}$O$^{+}$ 
observations is larger (by 50\%) than the 20$^{\prime\prime}$ range. \\
Some N$_{2}$ depletion is needed in order to explain the N$_{2}$H$^{+}$ and N$_{2}$D$^{+}$ 
observations. Through the hyperfine structure of these species we can determine directly the optical depth 
in several transitions using the relative intensities of the hyperfine satellites. This considerably reduces the 
error in our computations, compared to other species like HCO$^{+}$ and DCO$^{+}$.\\

This detailed model of the ion chemistry in L1544 simulates the observed depletion in the core center, and can 
reproduce the observed dependency of the column densities and abundances of species such as N$_{2}$H$^{+}$, 
N$_{2}$D$^{+}$, HCO$^{+}$, DCO$^{+}$ as a function of the impact parameter. This allows us to separately 
discuss the relative contributions from the high-density depleted core and the lower density foreground (and 
background) gas.

\subsection{Chemical parameter space exploration}
 
In order to focus and analyse in detail the H$_2$D$^+$ chemistry, 
we performed a parameter study using a model that computes the
deuterated forms of H$_3^+$ as a function of some key parameters, like
the grain size, the age of the L1544 condensation, and the cosmic rays
ionization rate.  This method has the advantage of concentrating on the 
H$_{2}$D$^{+}$ chemistry, avoiding to reproduce other molecular observations. 
Before discussing the comparison of the theoretical
predictions with the observations, we give a short description of the
model used. \\
It is an adaptation of the Ceccarelli \& Dominik
(2005) model, which has been developed for the proto-planetary disks.
It computes the H$_3^+$, H$_2$D$^+$, D$_2$H$^+$ and D$_3^+$ abundances
in cold and dense gas. Since the involved temperatures ($\leq 30$ K)
and densities ($\geq 10^5$ cm$^{-3}$) are very similar to those found
around L1544, the model can be used directly, by just changing the
geometry.  For an easy and straightforward comparison with the
observations we compute the H$_3^+$ chemistry in a gas cube with a
given density and temperature.  The relative abundances of the H$_3^+$
deuterated forms are computed solving the charge equilibrium equations
and the deuterium chemistry equations.  \\
In this model we also consider grains as a
possible source of H$_3^+$, H$_2$D$^+$, D$_2$H$^+$ and D$_3^+$
destruction.  In practice, the larger the CO depletion, the larger the
H$_2$D$^+$/H$_3^+$ and D$_2$H$^+$/H$_2$D$^+$ ratios.  \\
Two factors 
(other than the dust temperature) can modify the CO depletion : the age of the
condensation (larger ages give larger CO depletions because the
molecules have more time to freeze out onto the grains), and the gas
density (the condensation rate is proportional to the gas density). 
In addition, the cosmic ray ionization rate regulates the overall
ionization degree in the condensation, and therefore the H$_3^+$
isotopomers abundances. Finally, the grain sizes enter both in the CO
condensation rate (via the grains area), and in the charge balance,
because negatively charged grains recombine with the positively charged
molecular ions. In this model, we adopted the same parameters 
(binding energy for CO and N$_{2}$, sticking coefficient) 
chosen for our best-fit model (see section 4.1).\\

\subsubsection{H$_{2}$D$^{+}$ and D$_{2}$H$^{+}$ versus CO depletion}

\begin{figure}
\includegraphics[angle=0,scale=0.7]{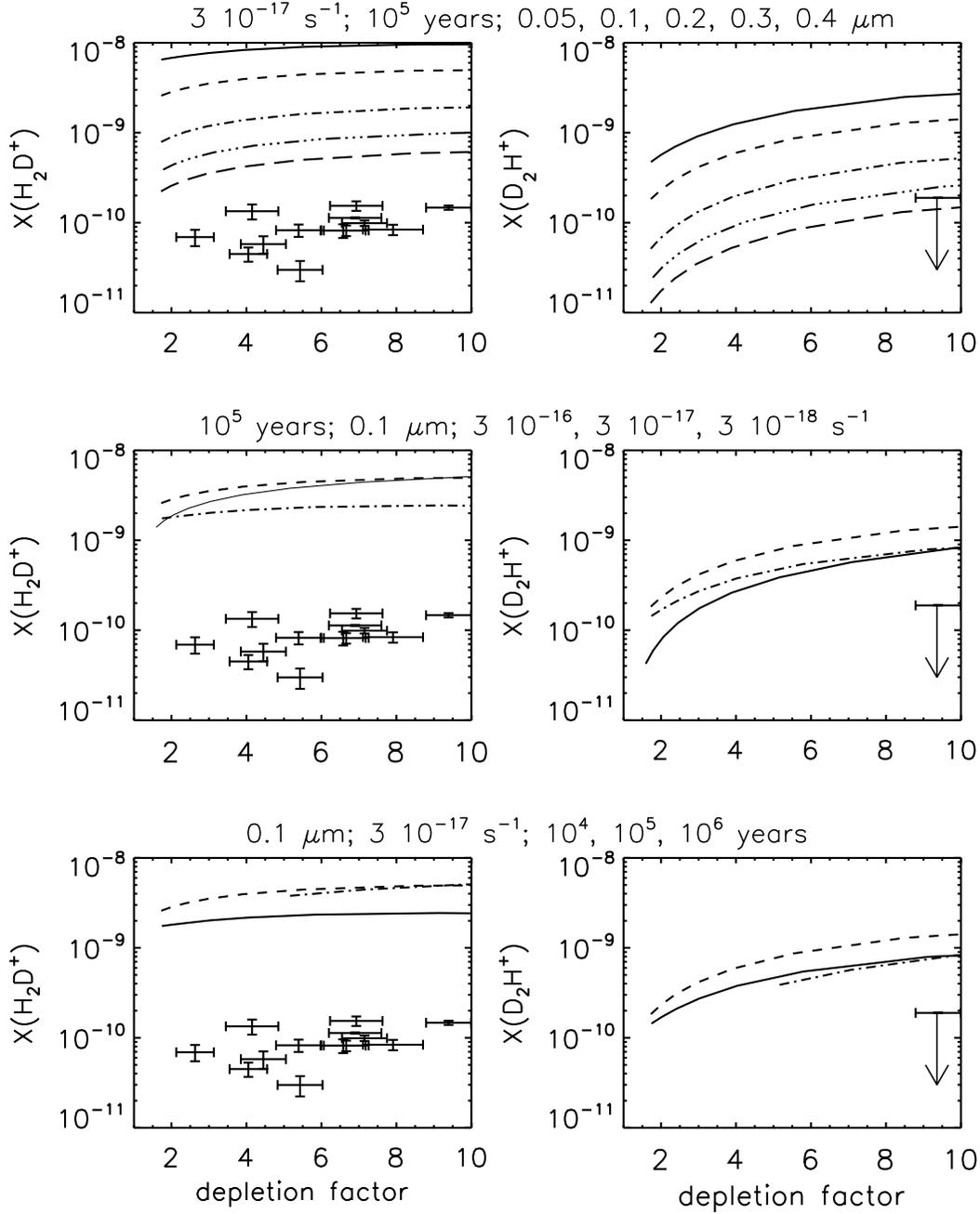}
\caption{Variation of the H$_{2}$D$^{+}$ abundance (respectively D$_{2}$H$^{+}$) for the model (plain line) as 
a function of CO depletion factor. The points with the corresponding error bars represent the observed abundances 
of ortho--H$_{2}$D$^{+}$ and para--D$_{2}$H$^{+}$ towards L1544 whereas the lines show the modeled abundances 
of the ortho + para states. Hence we expect the observed 
values to lie below the modeled ones. The temperature was fixed to 8 K. In the upper plot we varied the grain size: 0.05, 
0.1, 0.2, 0.3 and 0.4 $\mu$m. Note that increasing the grain size decreases the H$_{2}$D$^{+}$ and D$_{2}$H$^{+}$ abundances 
(see text). In the middle plot we varies the cosmic ionization ray: 3 10$^{-16}$ (plain line), 
3 10$^{-17}$ (dashed line) and 3 10$^{-18}$ s$^{-1}$ (dot-dashed line). In the lower plot we varied the age of the cloud: 10$^{4}$ 
(plain line), 10$^{5}$ (dashed line)and 10$^{6}$ years (dot-dashed line). \label{depletion}}
\end{figure}

In Figure \ref{depletion}, we present the results of the model, varying the three key parameters of the model 
(cosmic ionization rate, the grain radius and the age of the core) in order to reproduce the total (ortho + para) 
H$_{2}$D$^{+}$ and D$_{2}$H$^{+}$ abundances. We plot the abundances as a function of depletion because 
this parameter is more directly observed (via CO column density and dust continuum observations) than the gas 
density. We also make a comparison with the observations of 
ortho--H$_{2}$D$^{+}$ and para--D$_{2}$H$^{+}$ to get an insight into the deuterium chemistry of H$_3^+$.\\
We fixed the temperature of the cloud in the model to 8 K, which is within the range found from molecular and dust 
measurements. For comparison we also ran the cases with a temperature of 10 K and did not find any substantial difference.
The free-fall time is approximately 3 $\times$ 10$^{4}$ years for a density of about 10$^{6}$ cm$^{-3}$. In 
presence of a magnetic field, the collapse time is about an order of magnitude larger (Ciolek \& Basu 2000). 
We will vary the age between 10$^{4}$ to 10$^{6}$ years, the latter being close to the lifetime 
of a starless core. In our calculations, we assumed all the grains in the cloud to have the same size, but we investigated the 
result for different values of the grain radius. A grain size of 0.1 $\mu$m is the typical value assumed in 
chemical models for the interstellar medium, where it follows the MRN distribution. 
In the upper plot of Figure \ref{depletion} we used a standard cosmic ionisation rate of 
3 10$^{-17}$ s$^{-1}$, a typical age of 10$^{5}$ years and 
varied the dust grain averaged sizes between 0.05 $\mu$m and 0.2 $\mu$m. 
In the middle plot we fixed the age to 10$^{5}$ years, the grain size to 0.1 $\mu$m and varied the cosmic ionization 
rate between 3 10$^{-18}$ to 3 10$^{-16}$ s$^{-1}$. 
In the lower plot we fixed the grain size to 0.1 $\mu$m, the cosmic ionization rate to 3 10$^{-17}$ s$^{-1}$ and varied 
the age of the cloud between 10$^{4}$ and 10$^{6}$ years. 
The observation points (or upper limit in the case of D$_{2}$H$^{+}$) and their corresponding error bars are supperposed 
in these plots: ortho--H$_{2}$D$^{+}$ on the left side and para--D$_{2}$H$^{+}$ on the right side.

As the grain size increases while the total grain mass is conserved, the abundance of grains relative to H$_2$ 
decreases  $\propto a_\mathrm{grain}^{-3}$, and also the grain surface area per H$_2$ decreases 
$\propto a_\mathrm{grain}^{-1}$.  As this effect slows down the freezeout of CO, the same CO depletion 
is therefore reached either after a longer time, or at a higer density.  Since we keep the age constant, the effect 
of the density is observed in Figure \ref{depletion}: larger grain sizes correspond to higher densities, at 
which the overall degree of ionization is smaller.  Since H$_2$D$^+$ is the dominating ion, this is directly 
mirrored in the H$_2$D$^+$ abundance. Also, decreasing the cosmic ionization rate leads to a decrease in the
abundances. Indeed, H$_{3}$$^{+}$ ions (and consequently their deuterated forms) are formed by the 
ionization of H$_{2}$ due to cosmic rays.  And finally, increasing the age of the cloud will increase their 
abundances, as the CO depletion rate is time dependant. Consequently, for a more evolved cloud the same 
CO depletion is achieved for lower densities, corresponding to a higher degree of ionization, which is again 
directly reflected in the H$_2$D$^+$ and D$_2$H$^+$ abundances.

\subsubsection{The ortho and para forms}

Both H$_{2}$D$^{+}$ and D$_{2}$H$^{+}$ molecules have ortho and para forms, corresponding to the spin states 
of the protons (for H$_{2}$D$^{+}$) or deuterons (for D$_{2}$H$^{+}$). In order to compare the modeled abundances 
with the observations of one spin state only, it is critical to know the ortho--to--para ratio for these two molecules. 
Under LTE conditions, at temperature T, the relative populations of the lowest ortho (1$_{1,1}$) and para (0$_{0,0}$) 
levels of H$_{2}$D$^{+}$ would be:\\
\begin{equation}
{\rm \frac{n(1_{1,1})}{n(0_{0,0})}=9 \times exp(-\frac{86.4}{T})}
\end{equation}
and the relative populations of the lowest ortho (0$_{0,0}$) and para (1$_{0,1}$) levels of D$_{2}$H$^{+}$ would be:\\
\begin{equation}
{\rm \frac{n(1_{0,1})}{n(0_{0,0})}=\frac{9}{6} \times exp(-\frac{50.2}{T})}
\end{equation}
With these formulae, at 8 K, the H$_{2}$D$^{+}$ ortho--to--para ratio would be $\sim$ 1.8 10$^{-4}$ and the 
D$_{2}$H$^{+}$ para--to--ortho ratio would be $\sim$ 2.8 10$^{-3}$. The ortho form of H$_{2}$D$^{+}$ is 
produced mainly in reactions of the para form with ortho--H$_{2}$ (e.g. Gerlich, Herbst \&  Roueff 2002). 
Therefore, its high o/p ratio is attributable to the relatively high ortho--H$_{2}$ abundance as first noticed by 
\citet{pagani92} . Because the o/p ratio 
is not thermalized at the low temperature considered here, the o/p H$_{2}$D$^{+}$ ratio is not thermalized either. 
This can be illustrated in Flower, Pineau des For\^ets, Walmsley (2004) model where, at temperatures lower than 
10 K,  a hydrogen density of 2 $\times$ 10$^{6}$ cm$^{-3}$ and a grain size of 0.1 $\mu$m, the o/p--H$_{2}$D$^{+}$ 
reaches unity and the p/o--D$_{2}$H$^{+}$ value is about 0.1. Increasing the grain size will decrease the grain surface, 
leading to a decrease of the H$_{2}$ formation rate. Therefore, the H$_{2}$ ortho--to--para ratio will obsiously decrease, 
as well as the H$_{2}$D$^{+}$ ortho--to--para ratio.\\

From our observations we find that para--D$_{2}$H$^{+}$/ortho--H$_{2}$D$^{+}$ is less than 1.3 at the dust peak emission 
assumed to be at 8 K. In the prestellar core 16293E \citep{vastel04} we measured a para--D$_{2}$H$^{+}$/ortho--H$_{2}$D$^{+}$ 
of 0.75 for an excitation temperature of 10 K. 
We also computed the H$_{2}$D$^{+}$ ortho--to--para ratio and an upper limit on the D$_{2}$H$^{+}$ para--to--ortho 
ratio by comparing the observed abundances of ortho--H$_{2}$D$^{+}$ and para--D$_{2}$H$^{+}$ with the total (ortho + para) 
abundances of H$_{2}$D$^{+}$ as calculated using the model described in the previous section. 
In Table \ref{ortho_para}, the ortho--to--para ratio for H$_{2}$D$^{+}$ and the para--to--ortho ratio for D$_{2}$H$^{+}$ 
are quoted in order to reproduce the values obtained by the model for different sets of parameters, which are the cosmic 
ionization rate, the age of the core and the grain radius. We can directly compare our results on H$_{2}$D$^{+}$ with the 
Flower, Pineau des Forets, Walmsley (2004) model, even if their study assumes complete depletion (i.e. that 
CO abundance should be less than 10$^{-6}$). Indeed the abundance of both ortho and para spin states of H$_{2}$D$^{+}$ 
depends on the ortho and para forms of molecular hydrogen (through the proton-exchanging reaction of H$_{3}$$^{+}$ 
with H$_{2}$ followed by reaction 1) which does not vary as a function of depletion. On the contrary, the abundance 
of both para and ortho forms of D$_{2}$H$^{+}$ is determined by reactions with HD (produced on the grain surfaces) and 
will therefore depend on the core depletion. The Flower et al. model predicts H$_{2}$D$^{+}$ ortho--to--para ratios larger 
than the maximum value of 0.3 we observed for a temperature of 8 K and a H$_{2}$ density of 
2 $\times$ 10$^{6}$ cm$^{-3}$ (Pineau des Forets, {\it private communication}) spanning ranges up to 0.4 $\mu$m of 
the grain radius. As a consequence, since the H$_{2}$D$^{+}$ ortho--to--para ratio decreases as a 
function of the grain radius, it is likely that this should be larger than 0.3 $\mu$m. This depletion of small 
grains in this core is consistent with grain coagulation since ice condensation is not enough to increase the grain radius. 

\section{Conclusions and Perspectives}

In this paper we studied the prestellar core L1544, focusing on the H$_{2}$D$^{+}$ chemistry 
throughout the cloud. It is now widely accepted that the H$_{2}$D$^{+}$ molecule is the main tracer of the 
CO depleted prestellar cores, and we point out in this paper that the H$_{2}$D$^{+}$ emission is extended 
(over 60$^{\prime\prime}$), and an excellent tracer of the dust continuum, with a emitting radius 
of about 7,000 AU in the case of L1544. Hence, the line profile of this molecule would provide a crucial 
guide to the dynamical behavior of the high-density core. It is likely that the double-peak profile found 
in the central position, as well as positions around, is broadened by the central infall and is absorbed in the 
outer parts of the core \citep{vandertak05}. \\
We first used a model of the ion chemistry coupled with the physical structure of the core of L1544, 
including the deuterated isotopologues of the H$_{3}$$^{+}$ ion. This simulates the observed depletion 
and can approximatively reproduce the observed dependence of the column densities of species like 
N$_{2}$H$^{+}$, N$_{2}$D$^{+}$, HCO$^{+}$, DCO$^{+}$, H$_{2}$D$^{+}$ as a function of radius.\\
This study reveals a correlation between the ortho--H$_{2}$D$^{+}$ abundance and 1) the 
CO depletion factor, 2) the degree of deuteration in the HCO$^{+}$ molecule, 3) the degree of deuteration 
in the N$_{2}$H$^{+}$ molecule. H$_{2}$D$^{+}$ will survive longer, at higher density than 
N$_{2}$H$^{+}$ and N$_{2}$D$^{+}$.\\
We then used a simpler model focusing on the H$_{2}$D$^{+}$ chemistry where we did a wide parameter study. 
We discuss how the H$_{2}$D$^{+}$ and D$_{2}$H$^{+}$ abundances depend 
on the cosmic ionization rate, the age of the core, and the grain radius, by varying these parameters. 
It appears that the most important parameters to reproduce the observed values is the grain radius 
as small grains accelerate the freeze-out of CO and the observed values are consistent with a freeze-out rate 
dominated by larger grains. Therefore, we found that to reproduce the observations we need a larger grain 
radius of 0.3 $\mu$m.\\
This study can be considered as a springboard for observations to come, since the current 
observations are limited by the poor atmospheric transmission at the relevant frequencies. 
Table \ref{future} lists some of the major telescopes and interferometers that can be used for 
the study of H$_{2}$D$^{+}$ chemistry in prestellar cores, proto-planetary disks and protostars. 
Probably, D$_{3}$$^{+}$ can not be observable because enhanced D$_{3}$$^{+}$ abundance implies 
very cold and very dense regions. Since D$_{3}$$^{+}$ is a symmetric molecule, it does not have 
rotational transitions and does have its bending modes are in the near infrared. Therefore, these transitions 
will only be observable in absorption against a strong near infrared continuum. H$_{2}$D$^{+}$ and 
D$_{2}$H$^{+}$ are hence the only tracers of cold, dense phases of molecular clouds prior to star 
formation.

\acknowledgments

The authors thank the staff of the CSO telescope for their support. CV is grateful to Malcom Walmsley 
for fruitful discussions and to Laurent Pagani for useful comments. PC acknowledges support from the 
MIUR grant "Dust particles as factor of galactic evolution". This research has been supported by 
NSF grant AST-0540882 to the CSO.

Facilities:  \facility{CSO}.

\clearpage

\clearpage

\clearpage

\clearpage

\clearpage

\clearpage

\clearpage

\clearpage

\clearpage

\clearpage

\begin{table}
\caption[]{Lines parameters, and column densities of the ortho--H$_{2}$D$^{+}$ and para--D$_{2}$H$^{+}$ 
observations for an excitation temperature of 8 K. The 1 $\sigma$ errors are noted in parenthesis. The upper limit 
(3 $\sigma$) on the para--D$_{2}$H$^{+}$ column density has been determined using a linewidth of  
0.27 km~s$^{-1}$ (see section 3).
\label{line_parameters}}
\smallskip
\begin{tabular}{|c|c|c|c|c|c|c|c|}
\tableline
Position                      &  rms           &   V$_{LSR}$  & $\Delta v_{\rm res}$   &  $\Delta$v    &   $\int$T$_{mb}$$\Delta$v  & N(o-H$_{2}$D$^{+}$) & tau\\
($^{\prime\prime}$)  & (K)            & km~s$^{-1}$  & km~s$^{-1}$                & km~s$^{-1}$ & mK~km~s$^{-1}$                 &   cm$^{-2}$                 & \\  
\tableline
(0,0)                         & 0.11           &  7.277(0.016)    &     0.039       &   0.50(0.03)      &   495(33)  &  1.78(0.10) 10$^{13}$  &  0.58(0.05)\\
(20,0)                        & 0.12           &  7.319(0.045)   &      0.039      &   0.52(0.12)      &   255(37)  &  7.77(1.01) 10$^{12}$  &  0.24(0.04)\\
(0,-20)                       & 0.10           &  7.382(0.040)   &     0.039       &   0.59(0.09)      &   232(33)  &  6.92(0.90) 10$^{12}$  & 0.19(0.03)\\
(-20,0)                       & 0.13           &  7.269(0.033)   &     0.039       &   0.52(0.06)      &   293(40)  &  9.14(1.10) 10$^{12}$  & 0.29(0.05)\\
(-20,-20)                     & 0.11          &  7.238(0.023)   &     0.039       &   0.20(0.06)      &   98(20 )   &  2.99(0.58) 10$^{12}$  & 0.24(0.05)\\
(-40,0)                       & 0.13           &                       &     0.077       &                        &  $<$ 49     &$<$ 1.46 10$^{12}$      &         \\
(-40,20)                      & 0.12          &  7.294(0.035)   &     0.039       &   0.27(0.07)      &   115(26)   & 3.45(0.73) 10$^{12}$  & 0.21(0.05)\\
(-20,20)                      & 0.21          & 7.316(0.018)    &     0.039       &   0.37(0.04)      &   386(54)   & 1.28(0.15) 10$^{13}$  & 0.58(0.10)\\
(0,20)                        & 0.08           & 7.094(0.015)    &     0.039       &   0.43(0.04)      &   288(22)   & 9.27(0.62) 10$^{12}$  & 0.35(0.03)\\
(20,20)                       & 0.09          &  7.256(0.026)   &     0.039       &   0.29(0.08)      &   111(15)   & 3.30(0.57) 10$^{12}$  & 0.19(0.03)\\  
(0,40)                        & 0.11           & 7.300(0.024)    &     0.039       &   0.18(0.05)      &   75(20)     & 2.24(0.55) 10$^{12}$  & 0.20(0.03)\\
(-20,40)                      & 0.12           & 7.138(0.032)   &     0.039       &   0.45(0.09)      &   206(34)   & 6.24(0.94) 10$^{12}$  & 0.23(0.04)\\
(-40,40)                      & 0.10           &                       &    0.039       &                         & $<$ 27      & $<$ 4.52 10$^{12}$    &\\
(20,-20)                      & 0.14           & 7.378(0.040)    &    0.039       &   0.49(0.06)      &   224(42)   &  6.79(1.15) 10$^{12}$ & 0.23(0.04)\\
(20,40)                       & 0.12           &                        &    0.077       &                        & $<$ 46       & $<$ 1.27 10$^{12}$    & \\
(20,-40)                      & 0.11           & 7.436(0.054)    &    0.077       &   0.46(0.10)      &   215(45)   & 6.55(1.23) 10$^{12}$  & 0.23(0.04)\\
(40,-40)                      & 0.12           &                        &    0.077       &               & $<$46       & $<$ 1.27 10$^{12}$     & \\
\hline
\hline
Position                      &  rms           &   V$_{LSR}$   &    $\Delta v_{\rm res}$ &  $\Delta$v    &   $\int$T$_{mb}$$\Delta$v  & N(p-D$_{2}$H$^{+}$) & tau\\
\hline
(0,0)                         & 0.072           &  7.290(0.007)   &     0.021                       &    0.27            &   $<$ 103                               & $<$ 2.27 10$^{13}$     & $<$ 1.13\\
\hline
\end{tabular}
\end{table}
 
 \clearpage

\begin{deluxetable}{ccccc}
\tablecolumns{5}
\tablewidth{0pc}
\tablecaption{H$_{2}$D$^{+}$ ortho--to--para ratio and upper limit on the D$_{2}$H$^{+}$ para--to--ortho ratio, 
varying the cosmic ionization rate, the core age and the grain radius. \label{ortho_para}}
\tablehead{
\multicolumn{3}{c}{parameters} &    $\frac{o}{p}$ H$_{2}$D$^{+}$    &  $\frac{p}{o}$ D$_{2}$H$^{+}$ \\
\cline{1-3}
\colhead{cosmic ionization rate} & \colhead{age}     & \colhead{grain radius} &   \colhead{}   & \colhead{}\\
\colhead{(s$^{-1}$)}                    & \colhead{(years)}  & \colhead{($\mu$m)}    &   \colhead{}   & \colhead{}} 
\startdata
3 10$^{-17}$                               & 10$^{5}$        &              0.05             &  0.02                &   $<$ 1.07\\
3 10$^{-17}$                               & 10$^{5}$        &              0.1               &  0.03                &   $<$ 0.15\\
3 10$^{-17}$                               & 10$^{5}$        &              0.2               &  0.08                &   $<$ 0.55\\
3 10$^{-16}$                               & 10$^{5}$        &              0.1               &  0.03                &   $<$ 0.29\\
3 10$^{-18}$                               & 10$^{5}$        &              0.1               &  0.07                &  $<$ 0.29\\
3 10$^{-17}$                               & 10$^{4}$        &              0.1               & 0.07                 &   $<$ 0.29\\
3 10$^{-17}$                               & 10$^{6}$        &              0.1               & 0.03                 &   $<$ 0.29\\
3 10$^{-17}$                               & 10$^{5}$        &              0.3               & 0.17                 &   $<$ 0.63\\
3 10$^{-17}$                               & 10$^{5}$        &              0.4               & 0.32                 &   No solution\\
3 10$^{-18}$                               & 10$^{5}$        &              0.4               & 0.33                 &   $<$ 11.81\\
3 10$^{-17}$                               & 10$^{4}$        &              0.4               & 0.33                 &  $<$ 11.81\\
\enddata
\end{deluxetable}

\clearpage

\begin{deluxetable}{cccccccc}
\tablecolumns{8}
\tablewidth{0pc}
\rotate
\tabletypesize{\scriptsize}
\tablecaption{Current and future facilities for the chemistry of H$_{2}$D$^{+}$ and D$_{2}$H$^{+}$ in prestellar cores, proto-planetary disks 
and protostars.\label{future}}
\tablehead{
Name                 &   Aperture            &   Platform               & Available     &\multicolumn{2}{c}{H$_{2}$D$^{+}$}          & \multicolumn{2}{c}{D$_{2}$H$^{+}$}\\
\colhead{}        &  \colhead{}        &   \colhead{}          & \colhead{} &\colhead{1$_{1,0}$--1$_{1,1}$} & \colhead{1$_{0,1}$--0$_{0,0}$} & \colhead{1$_{1,0}$--1$_{0,1}$}  & \colhead{1$_{1,1}$--0$_{0,0}$}\\                          
\colhead{}        &  \colhead{}        &   \colhead{}          & \colhead{} &\colhead{(372.4 GHz)} & \colhead{(1.37 THz)} & \colhead{(691.7 GHz)}  & \colhead{(1.48 THz)}}                          
\startdata
CSO                  &  10.4 m                & Mauna Kea (USA)   & Y                & Y & N & Y & N\\
JCMT                &  15 m                   & Mauna Kea (USA)   & Y                & HARP B & N & Y & N \\
SOFIA               &   2.5 m                 &  Airborne (747)      &  2007           & N & Casimir                    & Casimir & Casimir \\
                         &                             &                             &                    &     & GREAT (CONDOR) &             & GREAT (CONDOR)\\                    
Herschel  (HIFI)  &   3.5 m                 &    Space (L2)         &  2007           & N & N & Y & Y\\
ALMA                & 50 $\times$ 12 m & Atacama (Chile)    &  2010           & Y & N & Y & N\\
APEX                 &   12 m                  & Atacama (Chile)    &  2005-2006   & Y & CONDOR & Y & CONDOR\\
\enddata
\end{deluxetable}
 
\end{document}